\documentclass[twocolumn, superscriptaddress, prl, amsmath, amssymb, aps, 10pt, floatfix]{revtex4-2}

\usepackage{comment}
\usepackage{graphicx}
\usepackage{dcolumn}
\usepackage{bm}

\usepackage{xcolor}

\begin{document}

\preprint{APS/123-QED}

\title{\textbf{Rolling at right angles: magnetic anisotropy enables dual-anisotropic active matter} 
}

\author{Eavan Fitzgerald}
 \affiliation{Institute of Science and Technology Austria, Am Campus 1, Klosterneuburg 3400, Austria}

\author{Cécile Clavaud}
 \affiliation{Institute of Science and Technology Austria, Am Campus 1, Klosterneuburg 3400, Austria}
\affiliation{
   IPR, Université de Rennes, Campus Beaulieu, Allée Jean Perrin Bâtiment 10B, 35042 Rennes, France\\
    }
\author{Debasish Das}
\affiliation{
Department of Mathematics and Statistics, University of Strathclyde,
Livingstone Tower, 26 Richmond Street, Glasgow G1 1XH, UK}

\author{Isaac C.D. Lenton}
\affiliation{Institute of Science and Technology Austria, Am Campus 1, Klosterneuburg 3400, Austria}

\author{Scott R. Waitukaitis}\email{Contact author: scott.waitukaitis@ist.ac.at}
\affiliation{Institute of Science and Technology Austria, Am Campus 1, Klosterneuburg 3400, Austria}

\date{\today}

\begin{abstract}
We report on an experimental active matter system with motion restricted to four cardinal directions. Our particles are magnetite-doped colloidal spheres driven by the Quincke electrorotational instability. The absence of a magnetic field ($|\bm{B}|=0$) leads to circular trajectories interspersed with short spontaneous runs. Intermediate fields ($|\bm{B}| \lesssim 20$ mT) linearize the motion along the axis perpendicular to $\bm{B}$. At high magnetic fields, we observe the surprising emergence of a second, distinct linearization along the axis parallel to $\bm{B}$. With numerical simulations, we show that this behavior can be explained by anisotropic magnetic susceptibility.
\end{abstract}

\maketitle

\section{\label{sec:intro} Introduction}
Driven by the ubiquity of systems that are active and inherently non-equilibrium in nature, the field of active matter has emerged as a vibrant area of research in the last few decades.  Loosely defined, active matter is composed of autonomous `agents', each capable of consuming energy to perform work and self-propel into motion \cite{marchetti_hydrodynamics_2013,bowick_symmetry_2022}. Remarkably, the violation of equilibrium at the agent level, combined with interactions between agents, can lead to the emergence of global phenomena, in particular flocking and coordinated collective motion \cite{vicsek_novel_1995, voituriez_spontaneous_2005, mccandlish_spontaneous_2012, zhang_collective_2010, liebchen_collective_2017, chardac_topology-driven_2021, bricard_emergence_2013, bricard_emergent_2015}.  

Typically the motion of individual agents in active systems can be considered isotropic, \textit{i.e.}, there is no preferred direction of travel. For instance, in biological systems comprised of \textit{Escherichia coli} or microtubule–molecular motors, agents perform isotropic random walks via run-and-tumble dynamics \cite{berg1973bacteria, zhang_collective_2010} or the stochastic binding and unbinding of proteins, which drives extensile sliding or `inching' \cite{ndlec1997self,sanchez2012spontaneous, sumino2012large}, respectively. In synthetic systems, \textit{e.g.}~light-activated colloidal surfers \cite{palacci2013living} or electric-field activated Quincke rollers \cite{bricard_emergence_2013, pradillo_quincke_2019}, rotational diffusion randomly reorientates particles to result in isotropy. In certain systems external influences can break symmetry and anisotropize motion, but in most cases this results in linearization along a single axis. Examples include chemotaxis \cite{liebchen2018synthetic, stark2018artificial}, phototaxis \cite{villa2019fuel} and magnetotaxis \cite{vincenti2019magnetotactic} in both biological and biomimetic systems, or otherwise directing motion through magnetic or electric forces in synthetic systems \cite{buttinoni_clustering_2013, yan2016reconfiguring, kaiser_flocking_2017, snezhko2009self, reyes_garza_magnetic_2023}. In principle, active matter anisotropized in more complex ways could lead to new ways of directing or controlling collective motion, but to our knowledge no experimental realization of this principle currently exists.

In this work, we introduce an active matter system that is dual-anisotropized, \textit{i.e.}, not along a single linear axis, but instead along two orthogonal axes. Our agents are magnetite-doped spherical particles which achieve motion via Quincke rotation, an electrohydrodynamic instability \cite{quincke_ueber_1896, jones_quincke_1984}. In the absence of an external magnetic field, the particles' motion is perfectly isotropic, as rotation is induced by a spontaneous symmetry breaking event. By introducing a magnetic field, we firstly break this isotropy with a linearization perpendicular to the field, which dominates at low to intermediate field strengths. However, as we increase the strength of the magnetic field, we observe the emergence of a second linearization \textit{parallel} to the field. Using numerical simulations, we show that this is caused by anisotropic magnetic susceptibility within each particle, leading to a complex dynamical system governing its convergence to either the ``perpendicular mode" or the ``parallel mode" (hereafter meaning either parallel or antiparallel motion). Furthermore, we demonstrate that the steady state dynamics of the perpendicular mode resemble a fixed point attractor, while those of the parallel mode exhibit limit-cycle-like behavior.

\begin{figure*}
    \includegraphics{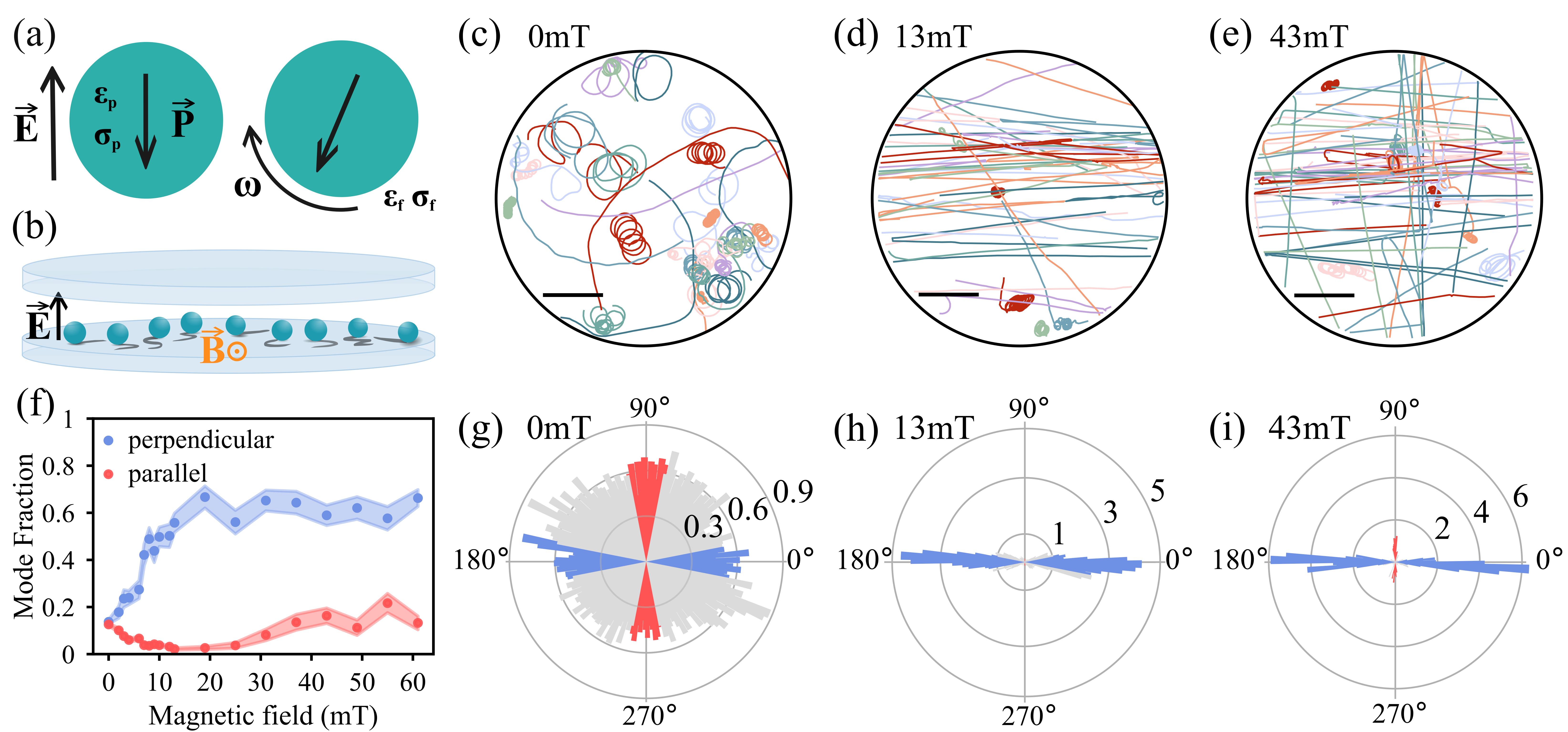}
    \caption{(a)~Intrinsically unstable dipole $\bm{P}$ induced in the particle by an applied electric field $\bm{E}$ when the condition \(\frac{\epsilon_f}{\sigma_f} < \frac{\epsilon_p}{\sigma_p} \) is met \cite{lampa1906dielectric}. For $\bm{E} > \bm{E}_c$ any small perturbation leads to persistent `Quincke' rotation, or rolling in the vicinity of a surface. (b)~Schematic of the experimental setup: particles lie on the bottom surface of our ITO-coated glass cell. A potential is applied between the glass electrodes to drive a DC electric field out of the plane of motion, while two electromagnets (not shown) drive a constant in-plane magnetic field $\bm{B}$. (c-e) The trajectories of an ensemble of rollers confined to a circular region (1 mm in diameter) at different magnetic field intensities and fixed potential $|\bm E| \approx 2E_c$ (scale bar is 200 \textmu m, see supplementary movies 1-3). The figures show a top-down view of the schematic shown in (b), with $\bm B$~aligned with the vertical axis and $\bm E$~coming out of the page. (c)~$\bm{B}$ = 0 mT: isotropic motion consisting of tight circular orbits interspersed with runs. (d)~$\bm{B}$~=~13~mT: first linearization of motion when a magnetic field is applied. A magnetic torque aligns the induced moment with the field, fixing the Quincke rotation axis to give rolling $\perp$ to $\bm{B}$. (e)~$\bm{B}$~=~43~mT: emergence of the second linearization, with motion parallel to $\bm{B}$ above 20 mT. The magnetic moment must have a non-zero component in the vertical plane, to drive motion collinear with $\bm{B}$. (f-h)~Normalized distributions of the angular displacements computed over 10-frame segments of the trajectories shown in (c-e). A 12\textdegree~tolerance window is defined and highlighted in blue (red) for characterization of the perpendicular (parallel) mode, with remaining displacements colored in gray. (h)~Homogeneous distribution in the absence of $\bm{B}$, corresponding to the trajectories in (c). (h)~Collapse in the distribution of angular displacements of trajectories shown in (d), as roller motion is driven perpendicular to the magnetic field axis when a 13 mT magnetic field is applied. (i) The emergence of dual-anisotropy with the coexistence of the perpendicular and parallel modes at higher field intensities, evident from the trajectories in (e). (f) The probability of observing a roller in a given mode at a specific field strength. This is estimated by computing the fraction of time each roller spends in the perpendicular or parallel mode, and the ensemble average of this fraction is plotted as a function of the magnetic field intensity.}
    \label{fig:1}
\end{figure*}

\section{\label{sec: exp results1} Rolling at right angles}

Quincke rollers have become an important model system for active matter due to their relative simplicity and rich dynamics \cite{karani_tuning_2019, pradillo_quincke_2019, pannacci_how_2007, bricard_emergence_2013, bricard_emergent_2015, kokot_spontaneous_2022, zhang_spontaneous_2023, maity_spontaneous_2023, mauleon-amieva_dynamics_2023, zhang_quincke_2021}. First observed by G. Quincke \cite{quincke_ueber_1896}, this electrorotational instability has since been well elucidated using a leaky dielectric model \cite{saville1997electrohydrodynamics, melcher1969electrohydrodynamics, jones_quincke_1984}, such that we will only briefly outline it here. A dielectric particle suspended in a weak electrolyte will have an anti-parallel dipole moment $\bm P$ when placed in an electric field $\bm E$~ as shown in Fig.~\ref{fig:1}(a). This anti-parallel alignment hinges on the ratio of the electrical permittivity $\epsilon_{f,p}$ and conductivity $\sigma_{f,p}$ of the fluid and particle respectively, which must fulfill the inequality \(\frac{\epsilon_f}{\sigma_f} < \frac{\epsilon_p}{\sigma_p} \). When this condition is met, and the electric field exceeds a critical value $E_c$, any perturbation to the dipole moment $\bm P$ drives the system towards steady rotation at a constant rate. At the critical field strength, the angular velocity $\omega$ undergoes a pitchfork bifurcation, reaching a steady state amplitude \(|\omega| = \tfrac{1}{\tau}\sqrt{(\tfrac{E}{E_c})^2 - 1}\), inversely proportional to the Maxwell-Wagner relaxation time $\tau = \tfrac{\epsilon_p + 2\epsilon_f}{\sigma_p + 2\sigma_f}$ and oriented in the plane perpendicular to $\bm E$. In the vicinity of a surface, rotation couples to friction, resulting in rolling at a fixed speed on the order of millimeters per second for a micron sized object. Given the stochastic nature of the perturbation which induces rolling (and the fact \(\bm{ \omega \cdot E} = 0\)), the initial direction of motion is random, rendering dilute ensemble- or time-averages isotropic.

Our setup is shown in Fig.~\ref{fig:1}(b). The particles are polymeric colloidal spheres doped with superparamagnetic nanoparticles---single magnetic domains---of iron-oxide (8~\textmu{m} diameter, $\sim$3-3.5\% magnetite content; UMDG003 COMPEL\texttrademark~from Bangs Laboratories). Nominally, the embedded nanoparticles render the spheres paramagnetic with essentially no remnant magnetization. We disperse the particles in an AOT-hexadecane solution (30~mM bis(2-ethylhexyl) sulfosuccinate sodium salt with 0.03\%~(w/v) H$_2$O, and a conductivity $\sigma_f\sim10^{-8}$ Sm$^{-1}$). The particles roll in a custom-made glass cell, confined between two ITO coated microscope slides (Sigma Aldrich, 75$\times$25 mm, 70-100 $\Omega$/sq). Circular or square conductive regions of varying sizes are defined by patterned deposition of SU-8 photoresist on the bottom slide via UV-lithography. Once the colloids have sedimented, a vertical electric field is applied between the electrodes to initiate Quincke rolling. A pair of electromagnets, with their cores oriented along the minor axis of the rectangular cell, generate a magnetic field in the plane of motion, which is nearly perfectly homogeneous across the domain. The rollers are observed from above with a Basler ace camera (acA2040) and recorded at framerates of 150-180~fps. Individual trajectories are extracted using the trackpy Python package \cite{allan_soft-mattertrackpy_2024} based on Crocker and Grier's \cite{crocker_methods_1996} tracking algorithm, and analyzed using custom scripts.

Observing the dynamics of our particles in the absence of a magnetic field yields trajectories characteristic of those in Fig.~\ref{fig:1}(c). These tight circular orbits, occasionally interspersed with `runs', are in contrast to the standard Quincke roller trajectories which are similarly isotropic for equivalent confinement, exhibit much larger persistence lengths. Previous studies using non-spherical active rollers have produced circular or chiral trajectories \cite{zhang_persistence_2021, kummel_circular_2013, chen_tunable_2023}, however they are much more ordered across the population and across experiments than what we observe in our system. While we suspect an underlying asymmetry to be behind these anomalous dynamics, in this work we simply wish to highlight the homogeneous distribution of the angular displacements in Fig.~\ref{fig:1}(g), in the absence of a magnetic field. Angular displacements (relative to the horizontal axis) over 10 frames are extracted from overlapping segments of the individual trajectories, and the normalized distribution of the displacements from all trajectories are plotted in Fig.~\ref{fig:1}(g-i).

\begin{figure*}
    \includegraphics{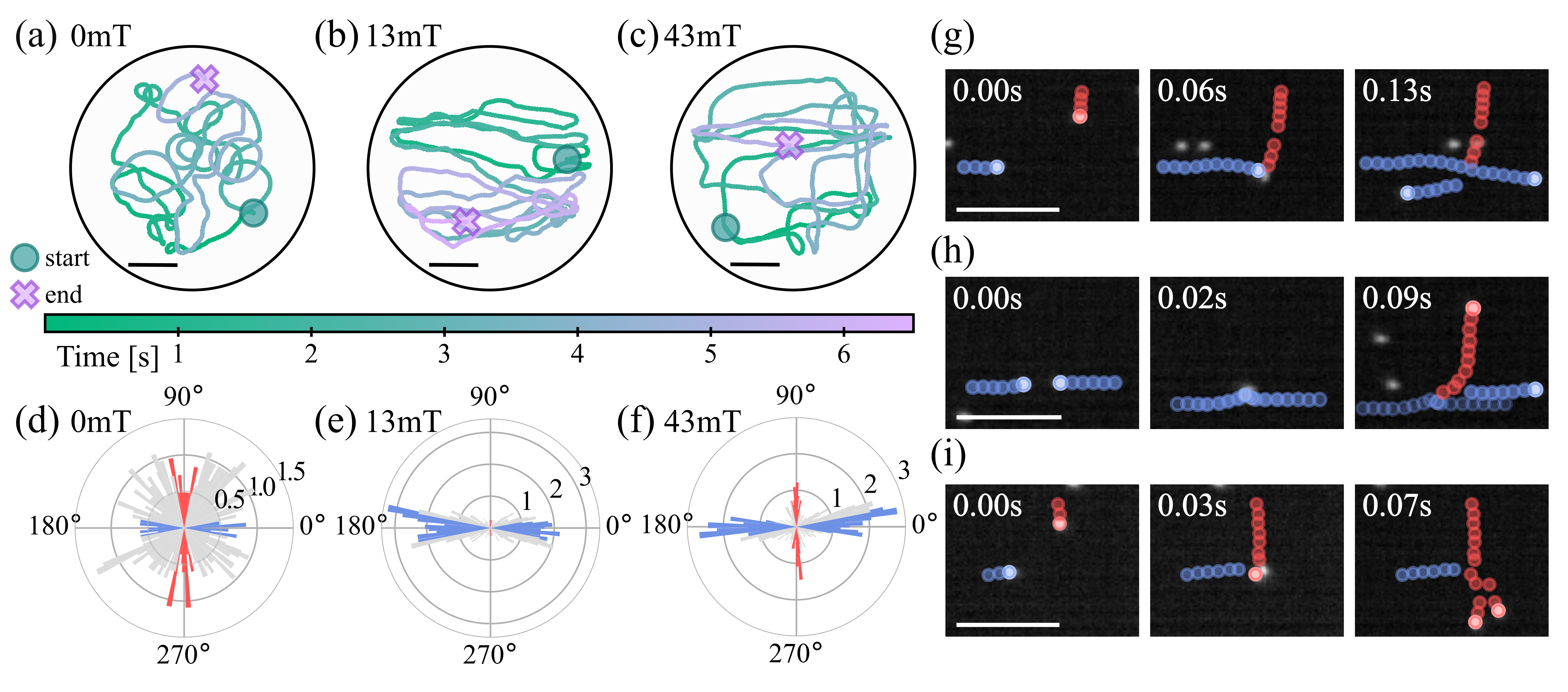}
    \caption{(a-c) The trajectory of a single roller confined to a 0.25 mm circular region (scale bar is 50 \textmu m, see supplementary movies 4-6), under the same field conditions as in Fig.~\ref{fig:1}, with $\bm B$~along the vertical axis and $\bm E$~coming out of the page. The trajectory is colored by time, with a circle (cross) denoting the start (end) point, while (d-f)~show the corresponding angular distributions. (a),~(d)~Circular motion and a homogeneous distribution at 0 mT. (b)~Motion perpendicular to the field dominates the trajectory, evident in the collapse of the angular distribution in (e)~at intermediate field strengths. (c),~(f)~Emergence of motion parallel to the field at higher field strengths, with the roller switching spontaneously between parallel and perpendicular modes. This points to the dual mode capacity of rollers, rather than an intrinsic preference for a particular mode at high fields. (g-i)~Stills from raw videos (see supplementary movie 7) capturing non-conservative interactions between rollers at $|\bm E| = 2E_c$ and $\bm B$~=~61~mT, demonstrate switching back and forth between perpendicular and parallel rolling. Sequential columns show the rollers before, during and after a collision respectively, as indicated by the annotated time. (g)~Two rollers, initially in orthogonal modes (perpendicular (blue) and parallel (red)), interact and both emerge in the perpendicular mode. (h)~A roller switches from perpendicular to parallel rolling, following a head-on collision with another perpendicular roller. (i)~A collision between orthogonal rollers triggers a switch from perpendicular to parallel rolling. We emphasize the two-way switching between modes, despite the perpendicular motion being more stable and energetically favorable. Scale bar is 50 \textmu m.}
    \label{fig:2}
\end{figure*}

What we focus on in the remainder of this work are the dynamics observed under the influence of a uniform magnetic field, with flux densities up to 62~mT. At intermediate field strengths (10-30~mT), we observe a striking linearization of the roller trajectories in the direction perpendicular to $\bm{B}$---the perpendicular mode---captured in Fig.~\ref{fig:1}(d). The coherence of the perpendicular mode is quantified in Fig.~\ref{fig:1}(g), with the collapse in the distribution of the angular displacements within a 12\textdegree~tolerance window, highlighted in blue. The saturation of this coherence around 15~mT is shown in the blue curve in Fig.~\ref{fig:1}(i). This curve represents the ensemble mean of the fraction of a trajectory spent in the perpendicular mode, but serves as an effective probability of observing perpendicular rolling as a function of the magnetic field strength.

As we increase the field strength beyond 20~mT and the saturation of the perpendicular mode, something quite unexpected happens. Unlike the uniaxial anisotropy also observed in recent work by Reyes-Garza \textit{et al.} \cite{reyes_garza_magnetic_2023}, we see the emergence of a second, distinct linear mode. A striking fraction of rollers execute stable paths parallel to $\bm{B}$, \textit{i.e.},~demonstrating simultaneous right-angled rolling. This dual-axis anisotropy is clearly evidenced in the trajectories at 43~mT in Fig.~\ref{fig:1}(e), and in the corresponding distribution of angular displacements in Fig.~\ref{fig:1}(h), where the displacements associated with the parallel mode are highlighted in red. The probability of observing parallel rolling as a function of the magnetic field strength is indicated by the red curve in Fig.~\ref{fig:1}(f), capturing its emergence once the field exceeds the saturation value of the perpendicular mode, and reaching a plateau around 40~mT.

A natural question that arises is whether certain rollers have an intrinsic preference for this parallel mode, or if all rollers have a dual mode capacity. To address this we refer to Fig.~\ref{fig:2}, firstly showing the dynamics of a single roller more tightly confined (domain size is 0.25 mm) than the previously discussed ensemble, but otherwise subject to the same experimental conditions. Fig.~\ref{fig:2}(a) shows the similarly isotropic motion when no magnetic field is applied, with the trajectory in this case colored by time. This isotropy is represented in Fig.~\ref{fig:2}(f) by the relatively homogeneous distribution of angular displacements. Introducing a magnetic field, the linearization of motion and dominance of the perpendicular mode at intermediate field strengths is duly replicated---see Fig.~\ref{fig:2}(b, e). Moving to higher field strengths, in Fig.~\ref{fig:2}(c, f) we finally observe right-angled rolling, clearly showing the bistability of orthogonal motion, with \ref{fig:2}(c) in particular illustrating a single roller switching back and forth between modes.

Returning to the ensemble case, the coexistence of both modes is also facilitated by interactions and collisions between rollers. In Fig.~\ref{fig:2}(g-i), we highlight three different cases of such collisions which demonstrate two-way switching between modes. Frames are extracted from raw videos and a time-lapse of the paths of colliding rollers are colored in blue or red, according to their respective mode. As can be seen in the figure, collisions between rollers may trigger a switch in its motional mode, but the result is not possible to predict \textit{a priori}. We do not observe a prevalence of alignment outcomes, as is typically exhibited by polar active matter via hydrodynamic effects \cite{marchetti_hydrodynamics_2013}. Hence, returning to the original question posed, we confirm the bistability of the modes for each roller, as demonstrated both by their individual capacity for right-angled rolling and, in the ensemble case, by observation of collision-mediated turns and the coexistence of parallel and perpendicular rolling.

\section{\label{sec:modelling} Modeling right angles}

Tackling the question of how this right-angled rolling emerges, and in particular how the parallel mode is sustained, begs an understanding of a roller's magnetic moment during motion. The first and most simple model we might consider is an ideal paramagnetic response, as bulk (\textit{i.e.}, not single-particle) data from the particle supplier would suggest. In this case, the induced moment of a roller would be $\bm m = \chi \bm B$, where the susceptibility, $\chi$, is a scalar constant. This model quickly collapses by the preclusion of a magnetic torque, since $\bm T_m = \bm m \times \bm B = \chi (\bm B \times \bm B) = 0$. Therefore, pure paramagnetism cannot account for any of our observations (\textit{i.e.}, the perpendicular or parallel modes).  

A second possibility to consider, and one that would sustain a steering magnetic torque, is the existence of a permanent dipole moment in each sphere. If this were the case, it is easy to see that the most energetically favorable state would have $\bm m \parallel \bm B$ at all times, with any deviation triggering a restoring magnetic torque. This enforces stable motion perpendicular to $\bm B$, fixed by $\bm m$ serving as an `axle' to the rolling sphere. This mechanism is consistent with Garza \textit{et al.}~\cite{reyes_garza_magnetic_2023}, who worked with 20~\textmu m~SiO$_2$ magnetic Quincke rollers and confirmed the alignment of the rotational axis using high-speed videography. However, while this simple model can explain the perpendicular mode, it cannot account for field-parallel motion. Qualitatively, persistent parallel motion requires an angular velocity (and hence a sustaining torque) orthogonal to the field. In such a configuration, $\bm m$ cannot function as a stable axle, as it occupies a torque-maximum orientation, and is continuously driven toward realignment with $\bm B$.  To maintain rolling parallel to the field, $\bm m$ must have a component directed out of the rolling plane. Hence, for $\bm m$ to be at least sometimes aligned with $\bm B$~while the sphere is in motion, it must undergo rotation in the vertical plane. However for a permanent moment, this necessarily means being driven towards anti-alignment with the $\bm B$ for half of its rotation cycle, a clearly unfavorable state. In reality, what occurs is a restoring torque which accelerates the sphere to stable motion perpendicular to $\bm B$, with $\bm m$~ stationary and aligned with the field. Hence, neither an intrinsic permanent magnetic moment, nor the paramagnetic-response model, can explain our dual-anisotropized system.

\begin{figure*}
    \includegraphics{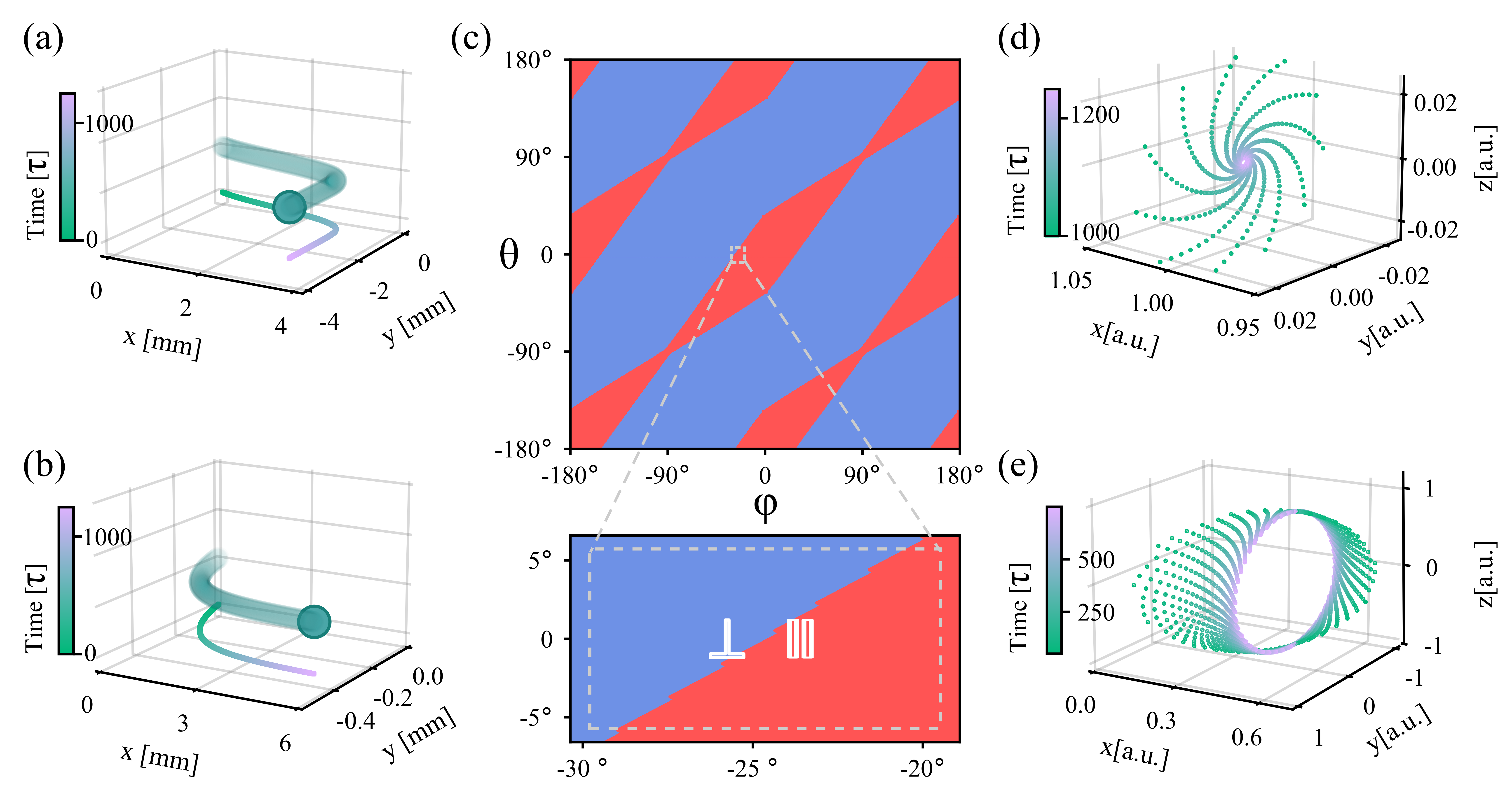}
    \caption{Results from dynamic numerical simulations of an isolated sphere with an anisotropic magnetic susceptibility under the influence of a driving electric and steering magnetic field. Simulations are carried out for a range of initial conditions given by $\phi$---the direction of motion---and $\theta$---the rotation of the soft axis in the xy-plane. (a) First 1250$\tau$ segment of a 5000$\tau$ trajectory resolving to perpendicular motion: $\phi$~=~25.5\textdegree, $\theta$~=~0\textdegree. (b) First 1250$\tau$ segment of a trajectory resolving to parallel motion: $\phi$~=~23.5\textdegree, $\theta$~=~0\textdegree. (c) Phase diagram showing the converged mode from a 5000$\tau$ sweep over pairwise values for $\phi$ and $\theta$ from -180\textdegree~to~180\textdegree, with resolution 400$\times$400. Blue (red) denotes the roller converging to the perpendicular (parallel) mode. Enlarged region shows the proximity of the initial conditions corresponding to the trajectories shown in (a, b): `$\bm \times$' (`$\parallel$') is the perpendicular (parallel) trajectory. (d) The dynamics of the soft axis as a function of time, in the frame of the virtual roller for the final 250$\tau$~of the trajectory in (a). The soft axis spirals towards a stable fixed point, located at the spiral center. Every 100$^{\text{th}}$ point is plotted, as for a  250~$\tau$~ segment at this scale, the spiral generated from the inclusion of every point would be overly dense. (e) The dynamics of the soft axis as function of time, in the frame of the roller body from $50-750\tau$~of the trajectory in (b), with every 100$^{\text{th}}$ point shown once again. The soft axis enters a steadily oscillating state, strongly reminiscent of a limit cycle in the xz-plane.}
    \label{fig:4}
\end{figure*}
In principle, to sustain perpendicular and parallel modes we need a body-fixed magnetic coupling that (a) leads to stable rolling when some aspect is aligned with $\bm{B}$, as in the permanent moment case, and (b) also permits a secondary stable mode when it is misaligned. These considerations lead us to explore a third possibility: anisotropic magnetic susceptibility. Mathematically, this corresponds to an induced magnetic moment of the form
\begin{equation}
    \bm{m} = \bm{X} \cdot \bm{B},
\end{equation}
where $\bm{X}$ is a $3 \times 3$ susceptibility tensor that rotates with the particle. This model differs from the ideal paramagnetic case in that $\bm{m}$ is not generally aligned with $\bm{B}$, allowing for sustained magnetic torques. It also differs from the permanent dipole model in that $\bm{m}$ is a linear response to the applied field, with magnetic energy $U = -\bm{B}^T \bm{X} \bm{B}$ that is always negative and bounded from below, eliminating the possibility of unstable anti-aligned configurations. It also permits a three-dimensional magnetic response that can stabilize the effective magnetic moment throughout rotation. In this way, anisotropic susceptibility enables field-responsive torques with complex orientational dependencies, potentially opening the door for richer dynamical effects.

We explore such effects using numerical simulations of an isolated Quincke roller with an anisotropic magnetic susceptibility. The leaky dielectric model explicitly describes the evolution of the electric dipole moment $\bm P$, while the magnetic susceptibility tensor $\bm X$ of our virtual sphere simply rotates with the angular velocity $\bm \omega$. The interaction between the components of the rotating moments not aligned with their respective driving field leads to sustained electric and magnetic torques acting on the roller, and determines the instantaneous angular velocity. Finally, in the absence of inertia, we balance electric and magnetic torque by the viscous torque acting on the roller yielding \cite{purcell_life_1977, jones_quincke_1984, turcu_electric_1987, bricard_emergence_2013, das_electrohydrodynamic_2013}
\begin{equation}
    \label{eq:torque_balance}
    4\pi\epsilon_f~\bm{P} \times \bm{E} + \frac{4\pi}{\mu}( \bm{X} \cdot \bm{B}) \times \bm{B} = 8 \pi \eta a^3 \bm{\omega} ,
\end{equation}
where $\mu$ is the magnetic permeability, $a$ is the radius of our spherical roller, $\eta$ is the fluid viscosity and all other symbols are unchanged. This leads to a closed system of equations that can be numerically integrated in time with a given set of initial conditions for the electric dipole and susceptibility tensor---see the Supplemental Material for more details \cite{sm} \nocite{goldman1967slow, beatty_finite_1986, gray_diffgeom_1997, weisstein_vivianis_nodate} . We consider a simplified case where the susceptibility tensor in its eigenbasis has just two elements, $\chi_s$ and $\chi_h$, such that the soft axis susceptibility $\chi_s > \chi_h$ defines the direction of greatest response. The tensor we consider can be written explicitly in its eigenbasis as
\begin{equation*} \label{Chi}
    \bm{X} = \begin{bmatrix} \chi_s & 0 & 0 \\ 0 & \chi_h & 0  \\ 0 & 0 & \chi_h \end{bmatrix},
\end{equation*}
and we set the ratio of the susceptibility coefficients to $\chi_s / \chi_h = 2$. We note that this ratio is merely chosen as a representative case, as a midpoint between perfect isotropy and uniaxial anisotropy, but is not a fixed property of the system, and does not qualitatively impact our results. We define the initial conditions of a simulated trajectory by the orientation $\phi$ of the velocity $\bm{v}$ relative to the magnetic field, and an initial orientation of the soft axis of $\bm X$ by some angle $\theta$ in the xy-plane (again relative to $\bm B$). We again refer the reader to the Supplemental Material \cite{sm} for additional information on the parameters and methods used.

Two example trajectories are presented in Fig.~\ref{fig:4}(a,~b), which immediately illustrate that anisotropic susceptibility is indeed capable of driving both perpendicular and parallel motion. Figure \ref{fig:4}(a) shows a sphere with initial conditions $\phi$~=~25.5\textdegree~and $\theta$~=~0\textdegree, which quickly turns to roll perpendicular to $\bm B $ in the steady state. By contrast, in Fig.~\ref{fig:4}(b) where the roller has initial conditions $\phi$~=~23.5\textdegree~and $\theta$~=~0\textdegree, the roller turns to demonstrate stable motion parallel to $\bm B$ in the steady state. These examples have initial $\phi$ values only 2\textdegree~apart, highlighting the sharp cut-off between initial conditions which result orthogonal steady states. To make this point even clearer, Fig.~\ref{fig:4}(c) presents a phase diagram for a full sweep of pairwise combinations of $\phi$ and $\theta$, where color indicates the resolved mode. The initial conditions of the trajectories shown in~\ref{fig:4}(a,~b) are plotted in the highlighted region. 

To gain insight into how this dual-anisotropized motion is permitted, in Fig.~\ref{fig:4}(d,~e) we plot the components of the soft axis in the frame of the roller as a function of time. Looking firstly at the case of perpendicular motion in Fig.~\ref{fig:4}(d), initially displaced from its $\theta$~= 0\textdegree~position, we see a transient regime of the soft axis spiraling inward as a function of time to ultimately align with the magnetic field. Hence, much like the case of the permanent moment, this mode corresponds to an energetic minimum of the magnetic energy at all times, where the soft axis is an `axle' parallel to $\bm B$. In the language of dynamical systems, the perpendicular mode can be considered a fixed point attractor, and all blue regions of the phase diagram in the Fig.~\ref{fig:4}(c) are its basins of attraction.

Shifting our focus to the case demonstrating parallel motion, we see that the soft axis resolves to an oscillatory state strongly reminiscent of a limit cycle. Fig.~\ref{fig:4}(e) shows the soft axis coordinates converging to a closed circular path in the xz-plane, traced by the 90\textdegree~phase-lag between $\chi_s^x$ and $\chi_s^z$, while $\chi_s^y$ oscillates with a comparatively negligible amplitude. This limit-cycle-like behavior is replicated in the dynamics of the induced magnetic moment, as well as the angular velocity, as $\omega_x$ and $\omega_z$ precess around the driving component $\omega_y$. This drives motion collinear with $\bm{B}$, but not perfectly so, as the roller `wobbles' back and forth over a single cycle. This characteristic `wobbling' is crucial for stability of the parallel mode, and a key distinction between the anisotropic susceptibility and permanent moment models. For the permanent moment, the sign of the magnetic torque is invariant under reflection about $\hat{\omega}$. Consequently, the direction of the angular velocity does not change over a rotation cycle, rendering the oscillatory dynamics associated with the parallel motion impossible. Conversely, the magnetic torque induced by a reflection of the tensor $\bm X$ about $\hat{\omega}$ in the anisotropy model is antisymmetric. Therefore as $\bm X$ co-rotates with the body of the roller, this antisymmetry triggers a change in the sign of the magnetic torque during a single cycle of rotation. This in turn leads to the periodic `wobbling' about $\hat{\omega}$, to ultimately sustain motion collinear with $\bm B$ and the oscillatory tumbling of the magnetic moment in the vertical plane, giving us the parallel mode. The result is oscillating steady state solution for the angular velocity, distinct from the steady fixed solution for the perpendicular mode---see the Supplemental Material and Figs.~S1,S2 for additional details \cite{sm}.

\section{\label{sec:discussion conclusion} Discussion and conclusions} 
A physical basis for the source of magnetic anisotropy in our particles is readily motivated by an inhomogeneous distribution of magnetite nanoparticles within the body of each roller, as well as a distribution of anisotropy axis orientations \cite{bean_superparamagnetism_1959, van_oene_biological_2015}. Moreover, as discussed in \cite{van_oene_biological_2015}, a clustering of nanoparticles which leads to dipole-dipole interactions invalidates the assumption of single isolated magnetic domains, culminating in a more complex magnetic response within the particle. Accordingly, the combined variability of these factors suggest a distribution of anisotropy within a population. In combination with other randomized or isotropic forces present in the absence of a magnetic field---and in particular those that cause the circular trajectories in Fig.~\ref{fig:1}(c)---this may account for the field-dependence of the modes in Fig.~\ref{fig:1}(i), as well as the observation that some rollers exhibit no response to the magnetic field whatsoever. Additionally, varying degrees of roller asphericity would not only contribute to an anisotropic magnetic response, but could also lead to the anomalous Quincke-rolling behavior observed when no magnetic field is applied.

Our results are consistent with a distinct class of active matter in which an anisotropic response is not constrained to a single axis. Most known polar systems are either isotropic or become anisotropic along a single axis when subject to external fields. With our `paramagnetic' colloidal particles and the Quincke instability, we observe striking dynamics consistent with a dually anisotropized system of two orthogonal linear modes. This behavior cannot be explained by either simple paramagnetism or fixed dipole moments. Instead, it emerges from an anisotropic paramagnetic response. The result is a rich single-particle phase space, where particles can spontaneously switch between modes, even under steady driving. These findings show that anisotropic properties at the agent level can expand the dynamical regimes accessible to active systems, and in doing so may open the door to unforeseen behaviors and new forms of control over both single-particle and collective dynamics.

\section{Acknowledgments}
This research was funded in whole or in part by the Austrian Science Fund (FWF) 10.55776/ESP298. For open access purposes, the author has applied a CC BY public copyright license to any author-accepted manuscript version arising from this submission. This project has received funding from the European Research Council (ERC) under the European Union’s Horizon 2020 research and innovation programme (Grant agreement No. 949120). This research was supported by the Scientific Service Units of The Institute of Science and Technology Austria (ISTA) through resources provided by the Miba Machine Shop, Nanofabrication Facility, Scientific Computing Facility, and Lab Support Facility. We wish to acknowledge the crucial contributions of  Alexandre Morin in getting the project off the ground, and Jack Merrin for creating the SU-8 deposition protocol used in the construction of our cells.

\bibliography{maintext, supplemental}

\end{document}


\preprint{APS/123-QED}

\title{\textbf{SUPPLEMENTAL MATERIAL \vspace{3 mm}\\ Rolling at right angles: magnetic anisotropy enables dual-anisotropic active matter} 
}

\author{Eavan Fitzgerald}
 \affiliation{Institute of Science and Technology Austria, Am Campus 1, Klosterneuburg 3400, Austria}

\author{Cécile Clavaud}
 \affiliation{Institute of Science and Technology Austria, Am Campus 1, Klosterneuburg 3400, Austria}
\affiliation{
   IPR, Université de Rennes, Campus Beaulieu, Allée Jean Perrin Bâtiment 10B, 35042 Rennes, France\\
    }
\author{Debasish Das}
\affiliation{
Department of Mathematics and Statistics, University of Strathclyde,
Livingstone Tower, 26 Richmond Street, Glasgow G1 1XH, UK}

\author{Isaac C.D. Lenton}
\affiliation{Institute of Science and Technology Austria, Am Campus 1, Klosterneuburg 3400, Austria}

\author{Scott R. Waitukaitis}\email{Contact author: scott.waitukaitis@ist.ac.at}
\affiliation{Institute of Science and Technology Austria, Am Campus 1, Klosterneuburg 3400, Austria}

\date{\today}
\maketitle
\renewcommand{\thefigure}{S\arabic{figure}}

\section{Experimental Methods and Materials}
\begin{flushleft}
\textit{Preparation of Colloidal Suspension---}The particles are purchased in an aqueous suspension, and therefore must be thoroughly washed and redispersed in the non-polar conducting fluid prior to use. A small sample of the aqueous suspension is firstly diluted ten-fold in de-ionized water. The following steps are then undertaken per round of washing for a 1 mL sample:
\begin{enumerate}
    \item centrifugation of the diluted suspension (604 $\times$~g rcf for 120 s), 
    \item removal of the supernatant using a 200 \textmu L pipette,
    \item resuspension of the particles (using a combination of vortex mixer and ultrasound) in fresh fluid.
\end{enumerate}\end{flushleft}
We use ethanol as an intermediate solvent before repeating the process for 8-10 rounds in the conducting fluid. The suspension can then be diluted as required before flowing into the cell. 
\begin{flushleft}
\textit{Magnetic Field Strength Calibration---}We calibrate the magnetic field strength $B$ across a range of driving currents $I$ (up to 0.5 A) using a PCE-MFM 3500 gauss meter prior to each experiment. Assuming the solenoid relation $B \propto I $, the field intensity listed for a given experiment is then inferred from the linear calibration curve.
\end{flushleft}

\begin{flushleft}

\textit{Computation of the Angular Displacements---}Trajectories are extracted from movies recorded at different magnetic field strengths using \textit{trackpy} and custom python scripts. Trajectories shorter than 100 frames are discarded, and individual trajectories are split into overlapping 10-frame segments, which is roughly equivalent to a distance of 8 body lengths for an 8 \textmu m diameter particle traveling at 1 mm/s. An angular displacement for each segment is calculated, to find a average velocity-orientation over this segment. In the main text, the computed angles are binned and plotted for all of the trajectories shown in Fig.1, and for the single particle's trajectories in Fig.2, at three sampled magnetic field strengths. 
\end{flushleft}

\begin{flushleft}

\textit{Quantifying the Modes---}
Fig.1(f) represents the fraction of time a particle spends traveling perpendicular or parallel to the field, as a function of the magnetic field strength. We estimate this ``mode fraction" by characterizing segments of the trajectory as perpendicular (parallel) mode if the angular displacement is within 12\textdegree of the horizontal (vertical) axis (for the layout indicated in the main text). 12\textdegree is chosen as a reasonable tolerance window, balancing resolution and statistical noise. The computed mode fractions for each trajectory are weighted by its duration relative to the ensemble. The ensemble mean is computed from these weighted values, and plotted in Fig.1(f), to represent the mode fraction as a function of field strength. Shaded areas represent the standard error of the mean for the ensemble.
\end{flushleft}

\section{Numerical simulations \label{sec:simulations}}
\begin{flushleft}
\textit{Setup and Parameters---}As described in the main text, we perform dynamic numerical simulations to probe the stability of an isolated virtual sphere with a fixed-body anisotropic magnetic susceptibility, traveling in a given direction relative to the axis of the magnetic field. We are particularly interested in whether our model can reproduce the dual-anisotropy we observe in our experiments, from a meaningful range of initial conditions. The direction of motion $\phi$ as well as the orientation $\theta$ of the soft axis, encoded in the anisotropy tensor $\bm X$, specify the initial conditions of a given simulation. Both $\phi$ and $\theta$ are defined in the plane of motion and relative to the magnetic field axis.\end{flushleft}

The instantaneous angular velocity $\omega$ is used to iteratively update the electric dipole moment $\bm P$ and the magnetic moment $\bm m$. The magnetic moment, unlike $\bm P$, is updated indirectly, instead $\bm X$ undergoes a rotation at each time step and $\bm m$ is calculated thereafter. We define the physical parameters used in our simulations on representative experimental or literature values. We use \(\epsilon_f = 2, \epsilon_p = 4\) for the electrical permittivity of the fluid and particle, denoted by the subscripts $f,p$ respectively, and conductivities \(\sigma_f \sim 10^{-8}~\text{S~m}^{-1}, \sigma_p \sim 10^{-14}~\text{S~m}^{-1}\). The conductivity of the fluid is measured experimentally, while the $\epsilon_{f,p}$ and $\sigma_p$ are inferred from the materials. These values define the characteristic Maxwell-Wagner relaxation time
\[\tau = \dfrac{\epsilon_p + 2\epsilon_f}{\sigma_p + 2\sigma_f} \approx 3.54 \text{ms}, \]
which is an important timescale representing the relaxation time of the electric dipole in response to a change in the driving electric field $E$. It is also useful to define the relative quantities $\epsilon_{pf}$, $\sigma_{pf}$
\[x_{pf} = \dfrac{x_p - x_f}{x_p + 2 x_f}, \]
where $x = \epsilon$~or $\sigma$, in specifying the dipole evolution equation 
\begin{equation}
    \dfrac{d\bm P}{dt} = \bm \omega \times [\bm P -a^3\epsilon_{pf}\bm E] - \frac{1}{\tau}[\bm P -a^3\sigma_{pf}\bm E],
\end{equation}
referred to in the main text, and to predict the critical value of the electric field $E_c$, which marks the onset of `Quincke' rolling. $E_c$ is determined not only by the electric properties of the system, but also by the viscosity of the fluid  $\eta$. Using literature values for the viscosity of hexadecane and lubrication theory to estimate the slipping coefficient $\beta$ \cite{goldman1967slow, pradillo_quincke_2019}, we obtain
\[ E_c \sim \beta \sqrt{ \dfrac{-2\eta}{ \tau \epsilon_f (\epsilon_{pf} + \sigma_{pf})}} \approx 1.5~ \text{MV m}
^{-1},  \]
where $\beta$ scales logarithmically with the lubrication layer thickness (10 - 100 nm according to Bricard \textit{et al.}\cite{bricard_emergence_2013}). This is quite close to our experimental observations, where we typically observe the onset of motion at $E_c^{exp} \approx 1.7$ MV m$^{-1}$.  

The susceptibility tensor $\bm X$ defined in the main text is characterized by two free parameters $\chi_s, \chi_h$ . We use $\chi_s/\chi_h$~=~2 for the results included, but as discussed in the main text, this is merely our chosen representative case. Indeed, additional simulations exploring generic $\bm X$ tensors have found that the coexistence of the perpendicular and parallel modes is an intrinsic feature of the anisotropic model, and not a result of a particularly special $\bm X$ we have chosen to present.

We use $E_c^{exp}$ to calculate an effective viscosity $\eta_{eff}$, which is in turn used to estimate two empirically grounded dimensionless quantities termed the Mason numbers. These describe the relative strengths of the field-driven and viscous torques, and are used to non-dimensionalize the physical quantities. The electric Mason number $Ma_E$ is given by 
\[Ma_E = \dfrac{\eta_{eff}}{\tau \epsilon_f E^2},\]
and similarly we can calculate the magnetic Mason number by
\[Ma_M = \dfrac{\eta_{eff}}{\tau \mu H^2},\]
where $ B = \mu H$, and $\mu$ is the magnetic permeability.  The angular velocity is derived from the torque balance equation in the main text
\begin{equation}
    \label{angular velocity}
    \bm\omega = \bm{P^*}\times \bm{E} + \bm{m^*} \times \bm{B},
\end{equation}
where $\bm P^*$, $\bm m^*$ are the dipole moments non-dimensionalized by the Mason numbers and a characteristic length scale, chosen to be the radius of our sphere $a$ = 4~\textmu{m}.  The ratio between the Mason numbers set the relative strength of the electric and magnetic torques. From our experimentally estimated values and chosen driving fields, we find
\[Ma_E / Ma_M \approx 0.009, \]
however stable perpendicular and parallel motion can be observed for a broad range of ratios obtained by scaling the physical quantities \textit{e.g.}~the magnetic field strength stipulated by $H$.

The simulation time step $dt$~is set to 0.005~$\tau$, where $\tau$~is the timescale of our system, \textit{i.e.},~the Maxwell-Wagner relaxation time. This time step was found to be sufficient to capture the dynamics of the system, and significantly shorter than the Maxwell-Wagner relaxation timescale, which is on the order of milliseconds. In standard units, $dt$~as well as the viscous timescale are on the order of \textmu{s}. The results presented in Fig.3 of the main text are from simulations with $10^6$~steps, or equivalently $\sim$17.7 s. For context, our experimental observation window is typically just a few seconds. The segment of the trajectory plotted in the main text corresponds to the first $2.5\cdot 10^5$ steps or $1250~\tau$, as the roller converges to either of the linear modes. The soft axis phase portraits are plotted for a shorter time frame, to highlight the characteristic dynamics of the fixed point attractor---spiral---or limit cycle---closed loop. The time frames shown are $1000-1250~\tau$ and $50-750~\tau$ in the perpendicular and parallel case respectively.

\begin{flushleft}
\textit{3D Rotation of the Susceptibility Tensor---}It is non-trivial to simulate a physically accurate rotation in three dimensions while avoiding numerical drift. We use the Euler-Rodrigues formulation \cite{beatty_finite_1986}, which characterizes a rotation by some magnitude (\textit{i.e.}~angle) around a given axis. In our simulations, the rotation matrix $\bm R$ is computed around the unit vector of the instantaneous angular velocity $\omega$, and the rotation angle $\delta$ is given by its magnitude scaled by the time step $dt$ \end{flushleft}
\begin{equation}
     \delta = ||\bm \omega ||~dt .\label{suppa} 
\end{equation}
We then construct the skew-symmetric cross-product matrix $\bm \Omega$ 
\begin{equation}
    \bm{\Omega} =  \dfrac{1 }{ ||\bm \omega ||}\begin{vmatrix} 0 & -\omega_z & \omega_y \\ \omega_z & 0 & -\omega_x  \\ -\omega_y & \omega_x & 0 \end{vmatrix}, \label{suppb} 
    \end{equation}
to subsequently compute $\bm R$ using the equation
\begin{equation}
    \bm R = \mathds{1}~+ \bm \Omega~\text{sin}~\delta + ( 1 - \text{cos}~\delta)~\bm{\Omega \cdot \Omega}, \label{suppc}
\end{equation}
 where $\mathds{1}$ is the identity matrix. This matrix encodes the counterclockwise rotation of $\bm X$~through the angle $\delta$ around~$\hat{\omega}$. Hence the rotated susceptibility matrix $\bm X'$ is simply
\begin{equation}
    \bm X' = \bm{R X R^T}.
\end{equation}
This formulation has the added benefit of implicit orthonormalization that prevents numerical drift during simulations with many time steps. We confirm this by checking the orthogonality of $\bm R$ via \( \mathds{1} = \bm{R^T R}\) and ensuring the determinant of $\bm R$ = 1 for the duration of the simulation.

\subsection{Additional Data from Simulations}
\begin{figure}[ht]
\includegraphics{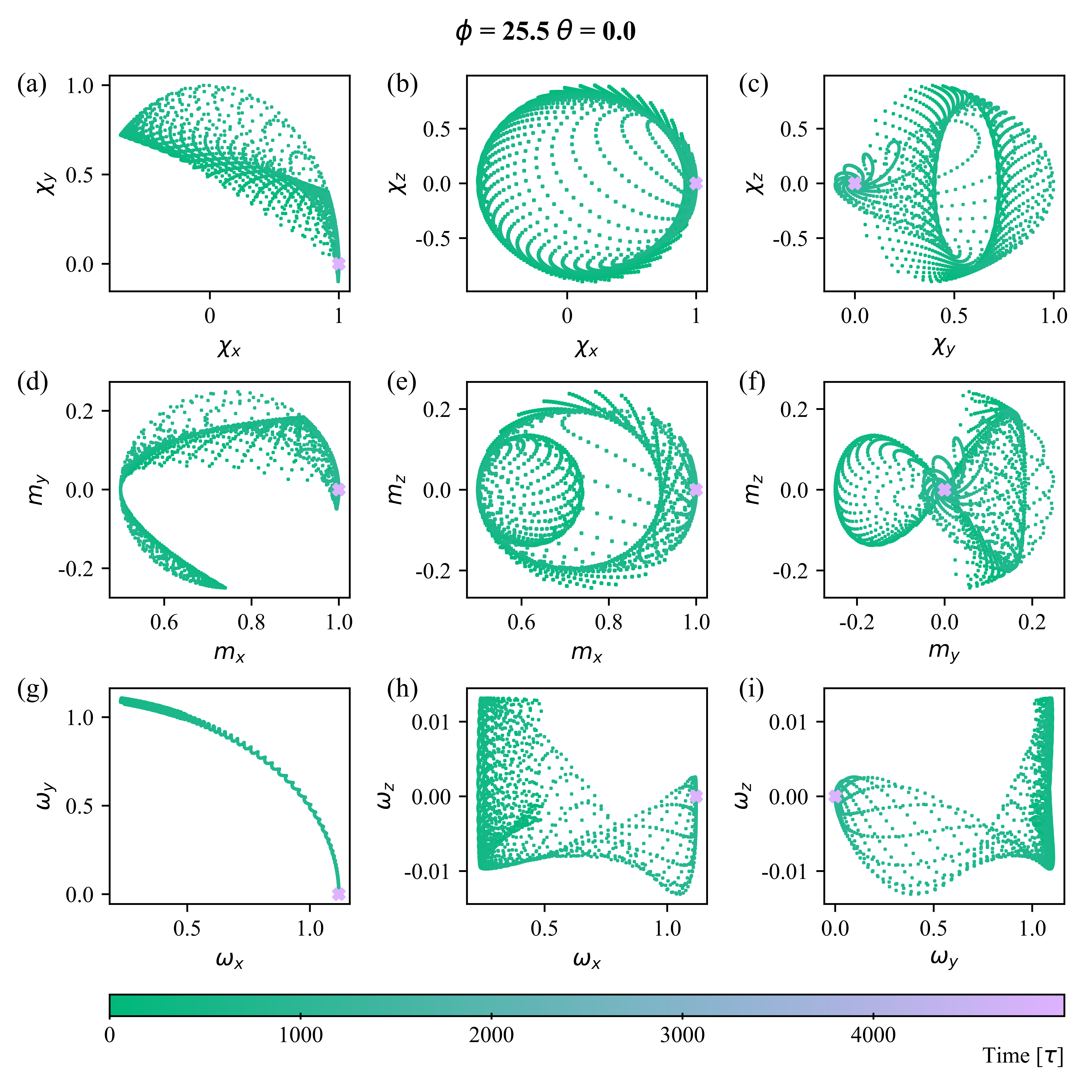}
\caption{Numerical simulation resolving to the perpendicular mode, as discussed in the main text. Runtime was 5000$\tau$ (equivalent to 17.7~s) with initial conditions $\phi~=~25.5$\textdegree, $\theta~=~0$. Every 100th data-point is plotted for ease of viewing. (a-c) are the cartesian components of the soft axis, (d-f) of the magnetic moment and (g-i) the angular velocity. Parameters converge to a fixed point once the roller is translating stably perpendicular to the field, with the final data-point plotted as an enlarged $\times$.}
\label{fig:SI fixed point}
\end{figure}

\begin{figure}[ht]
\includegraphics{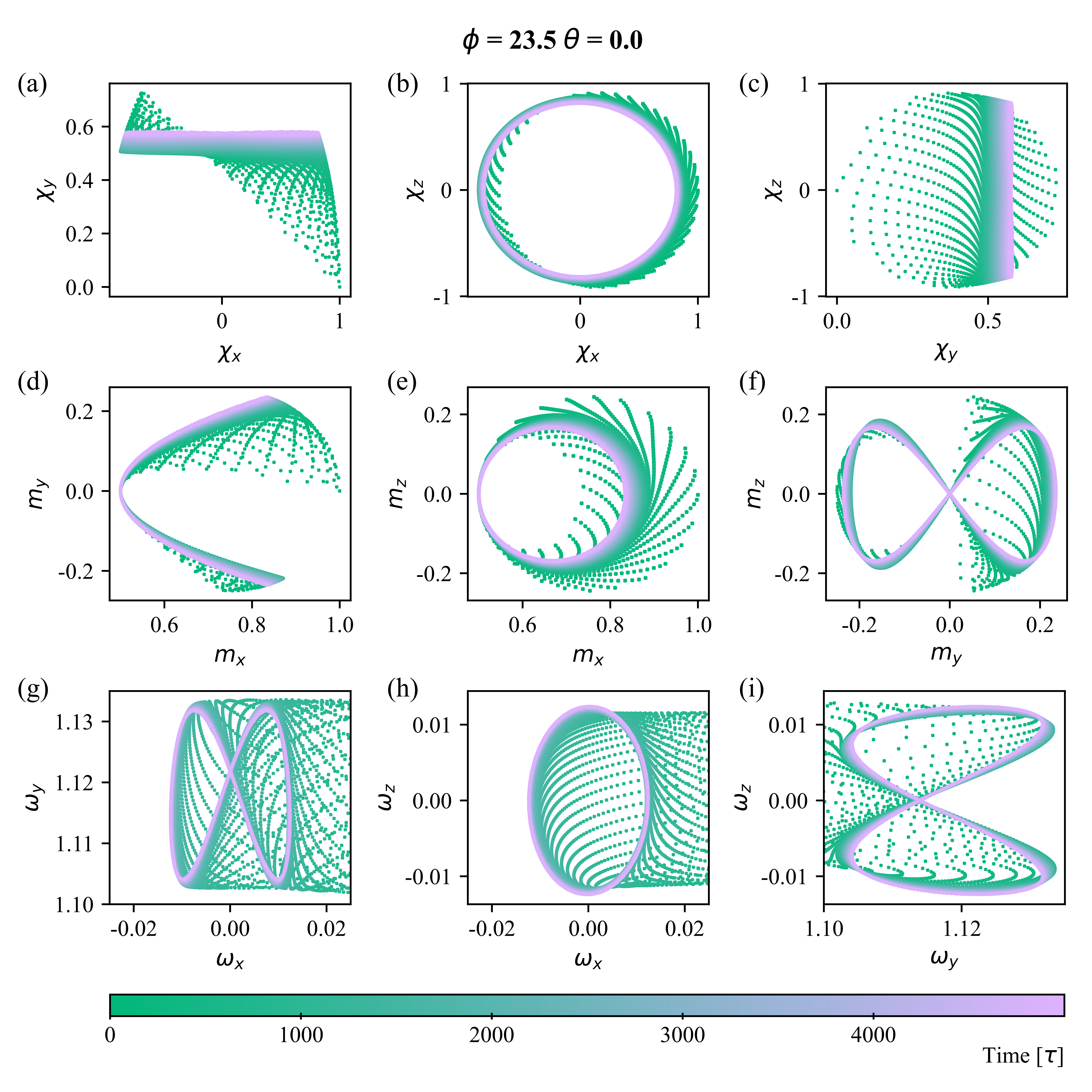}
\caption{Numerical simulation resolving to the parallel mode, as discussed in the main text. Runtime was 5000$\tau$ (equivalent to 17.7~s) with initial conditions $\phi~=~23.5$\textdegree, $\theta~=~0$. Every 100th data--point is plotted for ease of viewing. (a-c) are the cartesian components of the soft axis, (d-f) of the magnetic moment and (g-i) the angular velocity. Parameters converge to stable cyclical paths, strongly reminiscent of limit cycles. The field of view in (g-i) is cropped to focus on the oscillatory pattern of the angular velocity in the steady state, such that not all data points are shown.}
\label{fig:SI limit cycle}
\end{figure}

\begin{flushleft}
\textit{Dynamics of parameters in cartesian planes---}In Figs.~\ref{fig:SI fixed point}, \ref{fig:SI limit cycle} the cartesian components of the soft axis (a-c), the magnetic moment (d-f) and the angular velocity (g-i) are plotted for the trajectories discussed in the main text. All parameters converge to a fixed point in the case of perpendicular motion shown in Fig.~\ref{fig:SI fixed point}, appearing as an enlarged cross on each of the individual plots (intermediate times are stacked behind the final data point). The perpendicular mode is a stationary state and a fixed point attractor for the system, hence as the magnetic moment and the soft axis align with the magnetic field, the magnetic steering torque goes to zero resulting in a stationary solution for the angular velocity, driving motion perfectly perpendicular to the field. \end{flushleft}

In Fig.~\ref{fig:SI limit cycle} depicting the state with motion parallel to the field, the parameters resolve to a steadily oscillating solution strongly reminiscent of a limit cycle. In the case of the induced magnetic moment, the components appear to carve out a Viviani-like curve \cite{gray_diffgeom_1997, weisstein_vivianis_nodate}~with a lemniscate in the yz-plane, and closed orbit in the xz-plane. We note the swinging/tick-tock dynamics of $m_y$, as $m_x$ and $m_z$ steadily orbit one another, although the magnetic moment is shifted on the x-axis. This considerable alignment of the magnetic moment with $\bm B$ over the period of the limit cycle is crucial to the stability of parallel trajectories.

\begin{figure}[ht]
\includegraphics[width = 0.4\textwidth]{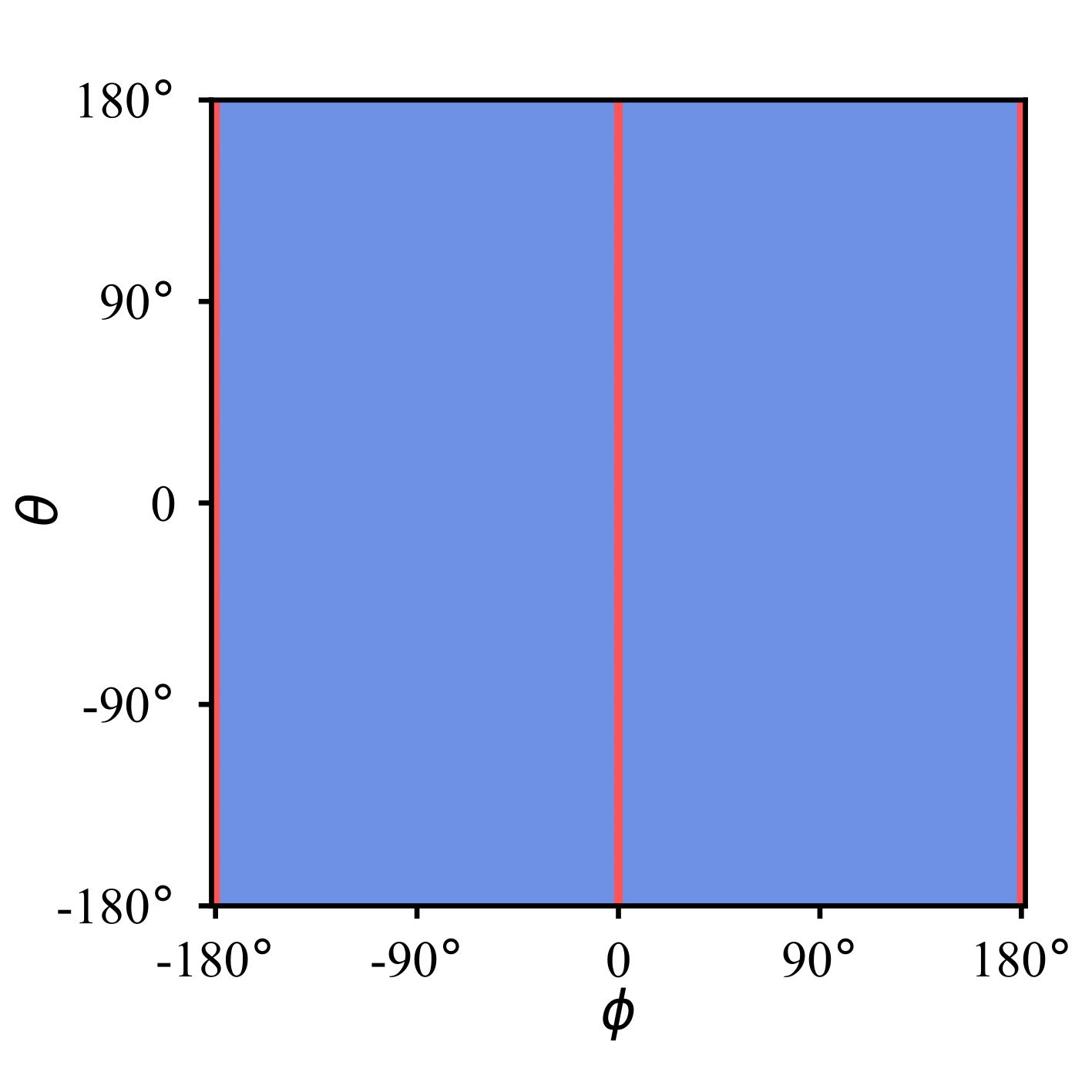}
\caption{Phase diagram showing the converged mode from a 5000~$\tau$ sweep over the initial conditions $\phi$ (direction of motion), and $\theta$ (rotation of the permanent moment in the xy-plane) with a resolution 400~$\times$~400. Blue (red) denotes the roller converging to a field-perpendicular (parallel) trajectory. In contrast to the phase diagram in Fig.3(c) of the main text, which uses the anisotropic susceptibility model, this plot clearly shows the inability of the permanent moment model to account for stable motion parallel to the field. Only for a highly idealized noise-free system which has an initial velocity oriented at precisely 0\textdegree~can motion parallel to the field persist for any period of time. All other conditions for $\phi$ result in a rapid convergence to motion perpendicular to the magnetic field.}
\label{fig:SI phase diagram}
\end{figure}

\begin{flushleft}
\textit{Permanent Moment Model---}Equivalent to the parameter sweep executed using the anisotropic susceptibility model, a sweep of simulations for all pairwise combinations of the initial conditions $\phi$ and $\theta$ was conducted for the permanent moment model, with the resultant mode  evaluated. In this setup, the magnetic dipole moment is directly updated at each time step, with no tensor rotation required---instead the moment itself, as a body-fixed axis, is directly rotated. The phase diagram generated from the converged state is plotted in Fig.~\ref{fig:SI phase diagram},  clearly showing the inability of the permanent moment model to account for stable parallel motion and our dual-anisotropized system. Only for an idealized noise-free case where the sphere has its motion directed precisely at $\phi$ = 0\textdegree~(or 180\textdegree), \textit{i.e.}~perfectly aligned to the magnetic field, can parallel trajectories persist for any length of time. All other conditions, even $\phi \sim$ 1\textdegree, rapidly converge to the perpendicular mode.\end{flushleft}

\section{Supplementary Movies}
\begin{flushleft}
\textbf{Movie S1.} Corresponds to Fig.1(c). Electric field was approximately $2E_c$ and no magnetic field was applied. Top view of a population of rollers recorded at 150 fps.

\textbf{Movie S2.} Corresponds to Fig.1(d). Electric field was approximately $2E_c$ and the magnetic field was 13 mT. Top view of a population of rollers recorded at 150 fps.

\textbf{Movie S3.} Corresponds to Fig.1(e). Electric field was approximately $2E_c$ and the magnetic field was 13 mT. Top view of a population of rollers recorded at 150 fps.

\textbf{Movie S4.} Corresponds to Fig.2(a). Electric field was approximately $2E_c$ and no magnetic field was applied. Top view of a single roller recorded at 180 fps.

\textbf{Movie S5.} Corresponds to Fig.2(b). Electric field was approximately $2E_c$ and the magnetic field was 13 mT. Top view of a single roller recorded at 180 fps.

\textbf{Movie S6.} Corresponds to Fig.2(c). Electric field was approximately $2E_c$ and the magnetic field was 43 mT. Top view of a single of roller recorded at 180 fps.

\textbf{Movie S7.} Corresponds to Fig.2(g-i). Electric field was approximately $E_c$ and the magnetic field was 61 mT. Top view of a population of rollers, recorded at 150 fps. Collisions are spliced from a longer movie, and shown with a cropped field of view. Interacting rollers are tracked and highlighted with a white circle, with their traces are shown to focus the viewer on the interaction. Playback is 5\% of original speed.
\end{flushleft}

\clearpage
\bibliography{supplemental, maintext}